# Magnetocrystalline anisotropy in YCo$_5$ and LaCo$_5$: A choice of correlation parameters and the relativistic effects

Manh Cuong Nguyen*, Yongxin Yao, Cai-Zhuang Wang, Kai-Ming Ho and Vladimir P. Antropov

Ames Laboratory – U.S. Department of Energy, Iowa State University, Ames, IA 50011, USA

*Abstract*

The dependence of the magnetocrystalline anisotropy energy (MAE) of MCo$_5$ (M = Y, La) on the Coulomb correlations and strength of spin orbit (SO) interaction within the GGA + U scheme is investigated. A range of parameters suitable for the satisfactory description of key magnetic properties is determined. The origin of MAE in these materials is mostly related to the large orbital moment anisotropy of Co atoms on the 2$c$ crystallographic site. Dependence of relativistic effects on Coulomb correlations, applicability of the second order perturbation theory for the description of MAE and effective screening of the SO interaction in these systems are discussed using a generalized virial theorem.

**PACS**: 75.30.Gw, 75.50.Vw, 71.15.Mb

The MCo$_5$ intermetallic compounds (M = rare-earth and Y) with the hexagonal CaCu$_5$-prototype structure have been attracted attention primarily due to their applications as permanent magnets. One of the key intrinsic magnetic properties determining such applications is the MAE. Essentially, all known magnets with the CaCu$_5$ structure exhibit a very high value of MAE with a maximum observed in SmCo$_5$ family of magnets, the most famous applied permanent magnets [1–3]. In general any MCo$_5$ system shows interesting properties, with even CeCo$_5$ demonstrating strong and temperature stable magnetic properties with different doping [4]. Theoretical studies of SmCo$_5$ are very limited due to difficulties in the description of localized 4$f$-states of Sm. In order to gain insights into the mechanism of high MAE in MCo$_5$, it is natural first of all to study the nature of MAE in similar systems where rare-earth atom is replaced by $d$-elements (Y or La) which do not have localized $f$-electrons but still demonstrate strong magnetic anisotropy. However, numerous studies of these systems using traditional band structure methods did not show a sufficient agreement with the experiment. Moreover, different methods produced very different results demonstrating a large sensitivity to the choice of the exchange and Coulomb correlations. A large value of anisotropy in a bulk metallic material can also be related to unusually strong relativistic effects in $f$-electron systems [3,5] and a question of the applicability of the traditional perturbation theory is naturally appearing. It would thus be desirable to understand how sensitive the important physics of these materials to both correlation and relativistic effects. A specific question which appears here is the influence of correlations on SO interaction that has been discussed recently [6].

There have been many theoretical works on YCo$_5$ and LaCo$_5$ systems to study the MAEs and other magnetic properties [7–15]. Density functional theory (DFT) calculations within local density approximation (LDA) gave different MAEs for YCo$_5$ depending on basis set and implementation. Some LDA calculations even predicted a wrong magnetization direction of YCo$_5$ [8]. DFT calculation within the generalized-gradient approximation (GGA) can predict a correct magnetization direction along the lattice vector **c** for YCo$_5$ but the obtained value of MAE was about a factor of 4 smaller than the experimental value [8]. The large anisotropy in YCo$_5$ and LaCo$_5$ is believed to be due to the large orbital magnetic moments of Co atoms as observed in experiments [7,16,17]. Some works have been performed taking into account the *ad hoc* orbital polarization (OP) correction to artificially increase the orbital magnetic moment [9,11,18,19]. Somewhat better values for MAE and orbital magnetic moments on Co atoms in YCo$_5$ have been obtained by these DFT + OP calculations in comparison with DFT [9,11]. However, the originally proposed OP correction due its unphysical atomic limit creates some uncontrolled anisotropies in YCo$_5$. A proper treatment requires implementation of a relativistic LDA+U type of scheme. There have been only a few calculations for MAE of LaCo$_5$ and they also showed very diverse results. A LDA + OP calculation by Steinbeck *et al* [9] showed a MAE of LaCo$_5$ ~ 9.0 meV/f.u., which is severely overestimated in comparison with the experiment value. While other LDA calculations showed a MAE of 0.38 meV/f.u. and 2.84 meV/f.u without and with OP correction, correspondingly [12].

In this work, we use standard DFT GGA + U scheme [20] taking SO interaction into account to investigate the magnetic properties of YCo$_5$ and LaCo$_5$. We show that there exists a "*universal*" set of U and J parameters which leads to a good agreement with the experimental data for the MAEs, orbital moment anisotropies (OMAs) and other magnetic properties for both YCo$_5$ and LaCo$_5$. Our calculation shows that the OMA is the key factor inducing the huge MAEs in YCo$_5$ and LaCo$_5$. We also verify the applicability of a second order perturbation theory for systems with large MAEs and discuss the screening of SO interaction by Coulomb correlations.

The spin-polarized DFT calculations are performed by the Vienna *Ab-initio* Simulation Package (VASP) [21] with a projector-augmented wave (PAW) pseudo-potential method [22,23] within the generalized-gradient approximation (GGA) parameterized by Perdew, Burke, and Ernzerhof [24]. The energy cutoff is 320 eV and the Monkhost-Pack scheme [25] is used for Brillouin zone sampling with a high quality k-point grid of $2\pi \times 1/60$ Å$^{-1}$, which is equivalent to a $14 \times 14 \times 15$ k-mesh. The total energy convergence criterion is $10^{-8}$ eV/unit cell. Strong-correlation effects are taken into account via a GGA + U scheme [20] for Co 3*d*-electrons. All calculations are performed with experimental lattice parameters from the ASM Alloy Phase Diagram Database. These parameters are a = 4.924, 5.105, and 2.698 Å, and c = 4.000, 3.966 and 3.710 Å for YCo$_5$, LaCo$_5$ and L1$_0$ CoPt, respectively. The MAE is calculated by force theory and is defined as the total energy difference for magnetization aligned along lattice **a** and **c**: K = E$^{[100]}$ – E$^{[001]}$. Here, E$^{[100]}$ and E$^{[001]}$ are the total energies with the magnetization aligned along the lattice **a** and **c** of the hexagonal or tetragonal unit cells, respectively. In details, a collinear self-consistent calculation is performed first. Then non-self-consistent calculations with SO interaction are performed with magnetization aligned along different directions, e.g. [001] and [100] to calculate $E^{[001]}$ and $E^{[100]}$, respectively. The SO anisotropy energy for an atom is defined in the same way: $K_{so} = E_{so}^{[100]} - E_{so}^{[001]}$. Here, $E_{so}^{[100]}$ and $E_{so}^{[001]}$ are the SO interaction energies with the magnetization aligned along the corresponding directions of that atom. The $K_{so}$ for the system is a sum of those from all atoms in the unit cell.

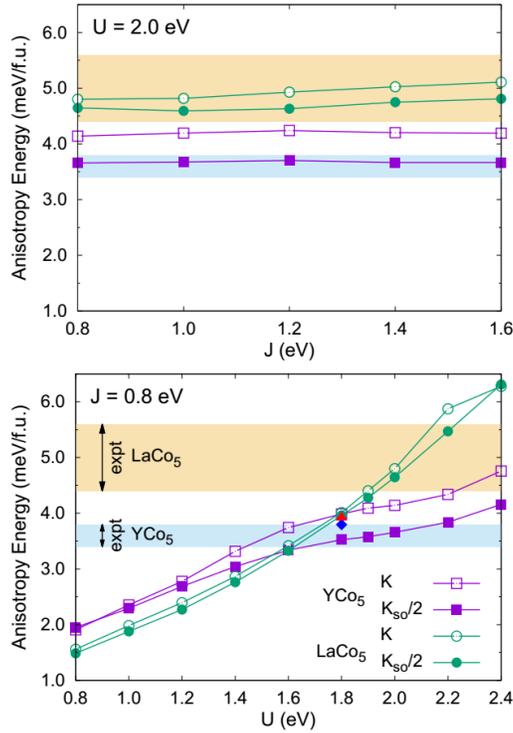

**Figure 1.** Dependences of YCo$_5$ and LaCo$_5$ MAEs on U and J parameters within a GGA + U scheme. Colored bands show the experimental range of MAEs. The blue diamond and red triangle are MAEs of YCo$_5$ and LaCo$_5$, respectively, and with U = J = 1.8 eV

First, we investigate the dependence of MAEs of YCo$_5$ and LaCo$_5$ on on-site Coulomb U and exchange J parameters as shown in Fig.1. Both Y and La does not have 4$f$-electrons and 4$d$-electron in Y and 5$d$-electron in La are not localized so the strong-correlation correction will be applied to Co 3$d$-electrons only. We vary U and J within reasonable ranges for 3$d$-transition metals: 1.0 to 2.4 eV for U and 0.8 to 1.6 eV for J, and with a step value of 0.2 eV. In Fig. 1, the golden and blue bands represent the range of MAEs for YCo$_5$ and LaCo$_5$ respectively reported by experimental measurements. The top panel of Fig. 1 shows the dependence of MAEs of YCo$_5$ and LaCo$_5$ on J when U is set to be 2.0 eV. As the changes in MAEs with J are very small, we use J=0.8 eV to investigate the dependence of MAE on the U parameter.

In contrast with a weak dependence on J, MAEs depend strongly on U. This can be seen in Fig. 1. For GGA + U with 1.0 eV ≤ U ≤ 2.4 eV and J = 0.8 eV, the MAE of YCo$_5$ varies from 2.40 to 4.70 meV/f.u. and it is in the experimentally observed range with 1.4 ≤ U ≤ 1.7 eV. For LaCo$_5$ the MAE value varies from 2.00 to 6.40 meV/f.u. and it is in the experimental observed range with 1.9 ≤ U ≤ 2.2 eV. It is interesting that for U from 1.8 to 2.0 eV, the MAEs of both $YCo_5$ and $LaCo_5$ are very close to the experimental values. The calculated MAE of $YCo_5$ varies from 3.99 to 4.14 meV/f.u. when U changes from 1.8 to 2.0 eV, about 4.8 to 8.9% overestimation in comparison with the upper bound of the experimental value. For the same range of U, the MAE of $LaCo_5$ is from 4.01 to 4.80 eV, about 8.4% smaller than the lower bound value to within the experimental value range. These results show that we can choose a common pair of on-site Coulomb U and exchange J parameters for Co 3d-electrons so that GGA + U can describe reasonably well the MAE of both $YCo_5$ and $LaCo_5$. This is very interesting: a "*universal*" pair of U and J for Co 3d-electrons can be chosen for metallic compounds with Co. We will show later that this choice of a universal set of U and J for Co also work very well for $L1_0$ CoPt, which is also an important permanent magnet but has a very different crystal structure from $YCo_5$ and $LaCo_5$. From the discussion above, a universal U can be chosen as any value between 1.8 to 2.0 eV. For all following calculations and analysis, we use an on-site Coulomb U = 1.9 eV and exchange J = 0.8 eV. It should be noted that the MAE of $YCo_5$ and $LaCo_5$ by GGA without U and J are 0.73 and 0.98 meV/f.u., respectively.

**Table I.** Atomic site resolved contributions to magnetic properties of $MCo_5$ (M = Y, La) and CoPt. MAEs from GGA + U calculation and experiment (K) are in meV/f.u., spin (S) and orbital (L) magnetic moments and OMA (ΔL) are in $\mu_B$/atom, $K_{so}$ is in meV/f.u.. $\Delta L^0$ and $K_{so}^0$ are $\Delta L$ and $K_{so}$ with U = J = 0 eV, respectively.

| System | | $YCo_5$ | $LaCo_5$ | CoPt |
|---|---|---|---|---|
| K | Calc. | 4.090 | 4.402 | 1.127 |
| | Expt. | 3.400 ÷ 3.800 | 4.400 ÷ 5.600 | ~ 1.000 |
| M/Pt | S/L | -0.419 / 0.037 | -0.283 / 0.027 | 0.334 / 0.068 |
| | $\Delta L/\Delta L^0$ | -0.020 /-0.009 | -0.014 /-0.009 | 0.010 / 0.019 |
| $Co_{2c}$ | S/L | 1.607 / 0.152 | 1.628 / 0.152 | |
| | $\Delta L/\Delta L^0$ | -0.062 /-0.020 | -0.058 /-0.027 | |
| $Co_{3g1}$ | S/L | 1.677 / 0.117 | 1.663 / 0.139 | 2.014 / 0.112 |
| | $\Delta L/\Delta L^0$ | -0.036 /-0.033 | -0.049 /-0.045 | -0.043 /-0.034 |
| $Co_{3g2}$ | S/L | 1.677 / 0.117 | 1.663 / 0.139 | |
| | $\Delta L/\Delta L^0$ | -0.020 / 0.003 | -0.032 /-0.008 | |
| M/Pt | $K_{so}/K_{so}^0$ | 0.447 / 0.073 | 0.957 / 0.213 | 2.189 / 1.945 |
| $Co_{2c}$ | $K_{so}/K_{so}^0$ | 2.106 / 0.250 | 1.795 / 0.186 | |
| $Co_{3g1}$ | $K_{so}/K_{so}^0$ | 0.867 / 0.622 | 1.335 / 0.983 | -0.109 /-0.781 |
| $Co_{3g2}$ | $K_{so}/K_{so}^0$ | 0.814 / 0.304 | 1.335 / 0.394 | |
| Total | $K_{so}/K_{so}^0$ | 7.156 / 1.803 | 8.553 / 2.355 | 2.080 / 1.164 |

Table I shows the atomic site resolved contributions to the magnetic properties of $YCo_5$ and $LaCo_5$. Note that when an external field is applied along lattice **a** and the SO interaction is included, the 3g-site of the crystal is split into 2 inequivalent sites. One site has a multiplicity of

1 (denoted as $3g_1$ hereafter) and the other has a multiplicity of 2 (denoted as $3g_2$ hereafter) [8,26]. The total spin magnetic moments of $YCo_5$ and $LaCo_5$ are 7.83 and 7.96 $\mu_B$/f.u., respectively. Taking orbital moment into account, the total magnetic moments are 8.52 and 8.71 $\mu_B$ for $YCo_5$ and $LaCo_5$, respectively. This agrees very well with the experimental observations of 8.33 $\mu_B$/f.u. for $YCo_5$ [7] and 8.46 $\mu_B$/f.u. for $LaCo_5$ [27]. From Table I, we can see that the orbital moment of Co is considerably larger than that of Y or La by about 3 to 4 times. This is consistent with experimental values. A recent X-ray magnetic circular dichroism (XMCD) experiment [15] showed that the orbital moments of Co on both sites in $YCo_5$ is about 0.2 $\mu_B$/atom with an orbital moment on the $2c$-site being slightly larger. By an inelastic spin flip neutron scattering experiment, Heidermann *et al* [28] showed that the orbital moments of Co on the $2c$ and $3g$-sites are 0.26 and 0.24 $\mu_B$ for $YCo_5$ and about 0.29 and 0.25 (by estimation) for $LaCo_5$. Our results for the orbital moment of Co in both $YCo_5$ and $LaCo_5$ are smaller than experimental values. The trend of larger orbital moment in $2c$-site is observed in our calculation. This trend of orbital moment is also consistent with previous calculations including an orbital polarization scheme [9,11] or an LDA + DMFT calculation [15] for $YCo_5$. We note that a LDA calculation with a linear-augmented plane-wave (LAPW) method [19] for $YCo_5$ observed the same orbital moments for Co on both the $2c$- and $3g$-sites.

From site resolved contributions to the SO anisotropy energy ($K_{so}$), we find that for $YCo_5$, the contribution from Co on the $2c$-site is ~2.5 times larger than that from Co on the $3g$-site and ~5.0 times larger than that from Y. These results agree with the experimental observation that the source of anisotropy of $YCo_5$ mainly comes from Co atoms with a dominant contribution from the $2c$-site [7,29]. Note that GGA calculations show that dominant contributions to MAE come from Co on the $3g$-site for both $YCo_5$ and $LaCo_5$. This is in contrast to our GGA + U calculation and the experimental results. The huge MAEs in $MCo_5$ (M = Y, La) compounds are ascribed to the large Co orbital moment and the large Co OMA [7,16,17,30]. In a model developed by Streever [17], the atomic anisotropy energy is proportional to the OMA. The OMAs of Co on $2c$- and $3g$-sites were measured from a polarized neutron scattering experiment [7]. Their values are -0.100 and -0.030 $\mu_B$/atom, respectively. This gives a total OMA ~ -0.30 $\mu_B$/f.u. for $YCo_5$, where the OMA is defined as the difference in orbital moment when the magnetization is aligned along lattice **a** or **c**. The OMAs from our calculation are -0.062, -0.036, and -0.020 $\mu_B$/atom for Co on $2c$-, $3g_1$, and $3g_2$-sites, respectively. The absolute values of OMAs from our calculation are somehow smaller than experimental values. They do, however, show the same trend that the $2c$-site Co OMA is much larger than that of $3g$-site Co by about 2 to 3 times. It is interesting that when we include the correlation effects, the OMAs are increased (Table I) and our results for OMA become similar to those obtained by LDA + OP [9], although the orbital moments of Co atoms in those calculations are about twice larger than ours. Note that the LDA + OP values for orbital moments of Co atoms are also larger than the values from flip flop neutron scattering [28] or recent XMCD [15] experiments. In addition, the MAE of $YCo_5$ from LDA + SO + OP is 4.40 meV/f.u., which is ~ 8% higher than the value from present calculation and ~ 16% higher than the upper bound of experimental value. The picture is very similar for LaCo$_5$ but the difference in OMA and K$_{so}$ between Co on 2$c$- and 3$g$-sites are smaller and the contribution from La to K$_{so}$ is larger in comparison with that from Y in YCo$_5$. These results and comparisons with experiments and previous calculations show that the GGA + U calculation with the universal pair of U and J can describe the magnetic properties of the MCo$_5$ compounds reasonably well.

In general, the model of Streever [17] is a simplified result of the second order perturbation theory. It has been known that the total SO anisotropy K$_{so}$ [3,31] can be presented as

$$K_{so} = (\lambda \Delta L_z + \lambda \Delta L_{\uparrow\downarrow})/2. \qquad (1)$$

Here, $\Delta L_z$ and $\Delta L_{\uparrow\downarrow}$ are anisotropies of the longitudinal and transversal components of the orbital moment operator. Our results indicate that the second (transversal) part is relatively small and does not play any significant role in the total magnetic anisotropy. The anisotropy of the longitudinal component (usual OMA) is dominating in Eq. (1). We emphasize that instead of the absolute value of orbital moment, a large OMA is a key factor in inducing huge MAEs in YCo$_5$ and LaCo$_5$. Overall, this implication is consistent with the physics of Streever's model [17] and is a very common behavior of anisotropy in metals. Our calculation also re-confirms that the main contributions to MAEs of YCo$_5$ and LaCo$_5$ are from Co on the 2$c$-site.

As an example demonstrating the optimal choice of U and J, we show in Table I the atomic resolved magnetic properties of L1$_0$ CoPt. CoPt was discovered to be a good permanent magnet material since the 1930s. However, the high cost of Pt metal has made this material a practically useless for permanent magnets applications. But CoPt is still of great interest for magnetic material and properties research. There have been many DFT calculations for MAE of CoPt. DFT calculations gave widely diverse MAE results depending on calculation and implementation of the computational method. This is similar to the situation of MCo$_5$ systems. For example, results from linear muffin-tin orbitals method are 2.29, 1.50, and 2.00 meV/f.u., and those from full-potential linear muffin-tin orbitals method are 1.05 and 2.20 meV/f.u. [32] within the LDA. Our LDA and GGA calculated MAEs are 1.53 and 0.84 meV/f.u., respectively. The experiment MAE of CoPt is ~ 1.00 meV/f.u. at T = 0 K and ~ 0.82 meV/f.u. at T = 300 K [32–34].

We perform a GGA + U calculation for CoPt with the universal U and J for Co 3$d$-electrons. Like in the calculations above, we do not apply additional correlation effects to Pt 5$d$-electrons because they are not localized. Our GGA + U result for MAE (1.127 meV/f.u.) agrees with the experiment. From Table 1 we also can see that the main contribution to MAE of CoPt is from the Pt atom. The Co atom shows a negative contribution (favoring in-plane magnetization or easy plane), but it is small in amplitude in comparison with the contribution from Pt. OMA is negative for Co and positive for Pt. These results are qualitatively consistent with recent work analyzing the constituents to magnetic anisotropy and other magnetic properties of CoPt [31,35]. Our value of MAE agrees well with experiments.

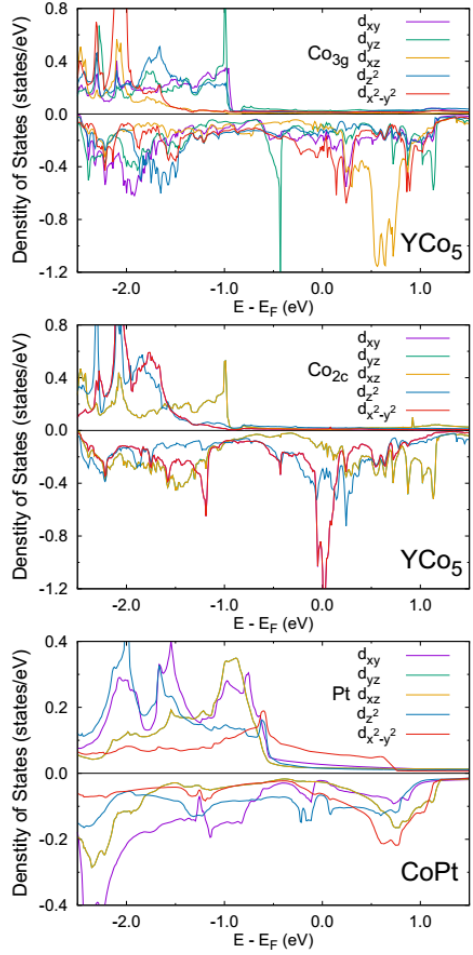

**Figure 2.** Projected density of states (PDOS) of Co in YCo$_5$ and Pt in CoPt from GGA + U calculations without SO interaction.

Figure 2 show the projected density of states (PDOS) on Co of YCo$_5$ and Pt of CoPt from GGA + U calculations without SO interaction. The positive and negative values of PDOS are the spin up and spin down components of PDOS, respectively. YCo$_5$ and LaCo$_5$ are isostructural and have very similar lattice parameters and the PDOS are very similar. We thus show only the PDOS from YCo$_5$. For YCo$_5$, the PDOS near the Fermi level are mainly contributed from the Co down spin, where the PDOS from the $d_{x2-y2}$-orbital of Co on the 2c-site are totally dominant. The PDOS of Y (not shown in plot) near the Fermi level is very small in comparison with that of Co. For CoPt, the spin down PDOS from different d-orbitals of Pt are comparable near the Fermi level. However, for the spin up channel, the PDOS from the $d_{x2-y2}$-orbital is very dominant and determines the opposite sign of OMA.

In all cases, the spin anisotropy of the DOS at the Fermi level is significant and allows a simple interpretation of large orbital moments in these systems if we use following relation [36,37]:

$$L = \lambda \sum m^2 \left( N_m^\uparrow(E_f) - N_{-m}^\downarrow(E_f) \right). \qquad (2)$$

Here, m is the magnetic quantum number and $N_m^{\uparrow/\downarrow}(E_f)$ are the spin up/down DOS at the Fermi level. Clearly, the largest contribution to the orbital moment is produced by states with a large m and the largest anisotropy of DOS. The relative signs of orbital moments obtained from eq.(2) and Fig.2 confirm a different behavior of OMA in CoPt and Y(La)Co$_5$ systems. The eq. (2) cannot be used for the calculation of OMA because it does not take into account the change of wave functions when SO coupling is included.

Let us discuss the relativistic effects and their dependence on Coulomb correlation from a general point of view. First, we stress that the total energy of a system when the atomic SO coupling $V_{so}$ is included is

$$E = T + V + V_{so}, \qquad (3)$$

and the resulting relativistic splitting is expected to be screened by a competition of kinetic and potential energies terms in eq. (3). Using second order perturbation theory, it has been shown (Ref. [31]) that the change in the total energy due to SO coupling addition is just half of the initial SO interaction energy

$$E^{(2)} = V_{so}/2. \qquad (4)$$

Correspondingly, the total anisotropy of hexagonal or tetragonal systems is K= $K_{so}$/2 [31]. This result is obtained in framework of traditional relativistic perturbation theory and is not necessarily valid in systems where a strength of SO coupling is comparable to the crystal field effects.

This result leads to an important statement: the strength of SO coupling in solids is always smaller ($\lambda = \lambda_0/2$) than the original (atomic) SO coupling $\lambda_0$ due to the "*screening*" reaction of the system and cannot be enhanced in this approach overall. We would like to emphasize that this screening is determined by both kinetic and potential energies' actions. To illustrate it in more detail, one can write an analog of the virial theorem with SO coupling included (mass-velocity and Darwin correction are ignored here for simplicity). Taking into account that SO coupling is a homogeneous function of the third order (in case of pure Coulomb potential), this virial theorem can be written as

$$2T + V + 3V_{so} = 0. \qquad (5)$$

Using eqs. (3) and (5), the resulting total energy can be presented in several alternative forms:

$$E = (V - V_{so})/2 = -T - 2V_{so} = (T + 2V)/3. \qquad (6)$$

While the virial theorem in a form of eq. (5) is a very approximate relation for the real interaction in the solid, the point is that there is always a general relation between the three terms in eq. (3) and the total energy in eq. (6) can always be presented as a function of only two terms.

If we now take into account the perturbative eq. (4), one can show that the total energy change when SO interaction is added can be presented as

$$E^{(2)} = \Delta V/4 = -\Delta T/5, \qquad (7)$$

where the symbol $\Delta$ stands for the difference of the corresponding potential/kinetic energies with and without SO coupling. The large relativistic change of potential energy $\Delta V = 2V_{so}$ is a generic result and is valid for both LDA and LDA+U treatment of potential energy terms. The discussion of the total renormalization of the SO coupling constant must evidently take into account the corresponding change of the kinetic energy term (with opposite sign) $\Delta T = -5V_{so}/2$. Therefore, the resulting "*screening*" of SO coupling (determined by a competition of large kinetic and potential energies) is relatively small and is expressed by eq. (4). The importance of the kinetic energy screening has often been ignored [6,38] and has led to fictitious enhancement of the SO coupling due to correlation effects (see $\Delta V = 2V_{so}$ above). In addition, one can write the corresponding expressions for the kinetic and potential magnetic anisotropies analogous to eqs. (4)-(7).

Overall, this connection between different energy terms leads to an important conclusion: the total magnetic anisotropy of a system can be presented equally well using only kinetic, potential, or SO energy anisotropy. We also mention that with the addition of Hubbard terms, the total potential and its radial derivative are modified. Thus, the atomic (non-renormalized) $\lambda$ is also modified. In the majority of DFT + U formalisms, the Hubbard corrections are added without radial dependence and $\lambda$ is not changed. However, the SO energy is modified as new wave functions are affected by Hubbard potential terms. Now we will show how significant this modification becomes.

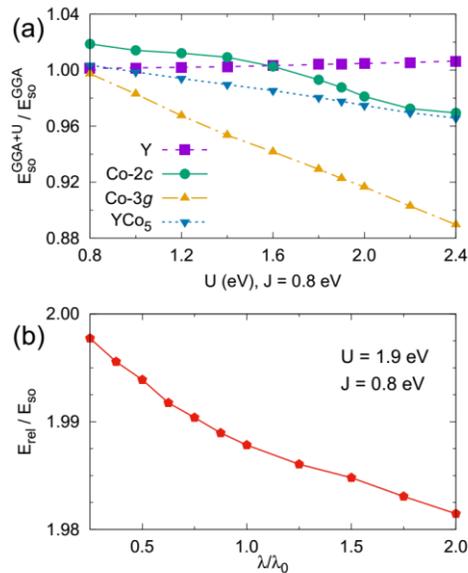

**Figure 3**. Ratios of (a) GGA + U and GGA SO energies as function of U and (b) total relativistic and SO energies as function of SO coupling strength within GGA + U for YCo$_5$.

We artificially vary the strength of Coulomb correlations (potential energy) and according to eqs. (3)-(7) this leads to the corresponding changes in the kinetic energy and overall screening of SO interaction. In our calculations, we did not see the enhancement of atomic SO coupling constants due to r-independent Hubbard terms addition. This was as expected. However, the Coulomb correlations have effects on the SO energy via the self-consistent wavefunction. On Fig. 3(a), the site resolved ratios of SO atomic energies from GGA + U and GGA calculations for $YCo_5$ are shown. The SO energy on all atoms is practically unchanged (within 4%) with correlations added and only the Co atoms on the 3*g*-site losing nearly 10% of the original SO energy. Fig. 3(b) shows the ratio of total relativistic and SO energies obtained in GGA + U. The ratio is almost two for the whole range of scaling SO coupling strength $\lambda/\lambda_0$, thus confirming that the SO coupling in this system is screened by half. This is due to both kinetic and potential energy competition as discussed above.

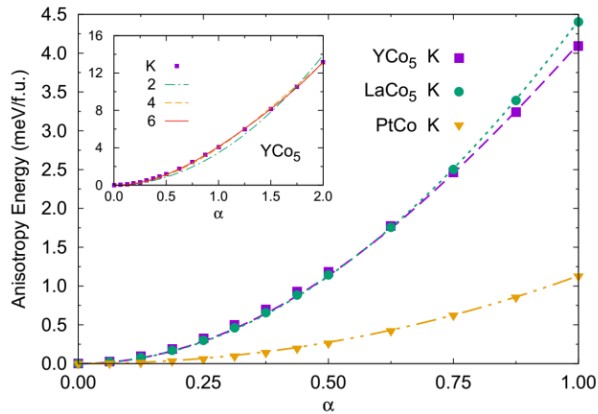

**Figure 4.** MAE as function of SO coupling strength for $YCo_5$, $LaCo_5$, and CoPt. Data points are from GGA + U calculations and the dashed and solid lines are fitting curves. Inset shows different fitting curves to MAE of $YCo_5$.

In Fig. 1, we also show the $K_{so}/2$ together with K as functions of U and J parameters in order to further verify the perturbation theory in treating SO interaction. Clearly, the $K_{so}/K$ ratio in $LaCo_5$ is almost two for different values of U. The deviation from the perturbation theory is larger for $YCo_5$ where $K_{so}/K$ is between 1.75 and 1.95. Another way to verify the perturbation theory is to investigate the change of MAE with SO coupling strength ($\lambda$). We scale the SO coupling strength by the factor $\alpha$, which ranges from 0 to 2. This corresponds to SO interaction that was previously not included to doubly strengthen. We extend $\alpha$ to 2 instead of 1 to get a better fitting which is able to predict the MAE if we can enhance the SO coupling. The results from these scaled SO coupling calculations are shown in Fig. 4. If the second order perturbation theory is obeyed, the MAE vs. $\alpha$ curve should be ideally parabolic: $K_2(\alpha) = k_2\alpha^2$. In our calculation for all $YCo_5$, $LaCo_5$ and CoPt, the MAE vs. $\alpha$ curves are fitted better to the function including not only the 2$^{nd}$ but also the 4$^{th}$ order terms [$K_4(\alpha) = k_2\alpha^2 + k_4\alpha^4$], or the function including 2$^{nd}$, 4$^{th}$, and 6$^{th}$ order terms [$K_6(\alpha) = k_2\alpha^2 + k_4\alpha^4 + k_6\alpha^6$], if we fit to the whole range of $\alpha$ from 0 to 2. The MAE vs. $\alpha$ curves can only be fitted very well to the parabolic $K_2(\alpha)$ function if we limit the fitting to small $\alpha$, i.e., smaller than 0.5. The $K_2(\alpha)$ function with fitted $k_2$ parameters describes the MAE with $\alpha > 0.5$ very poorly.

The inset of Fig. 4 shows the MAE from GGA + U calculations and 3 fitting functions: $K_2(\alpha)$, $K_4(\alpha)$, and $K_6(\alpha)$ for YCo$_5$. We can see that $K_2(\alpha)$ fits quite poorly to the MAE curve, while $K_4(\alpha)$ and $K_6(\alpha)$ fit very well to the MAE curve. The fitted parameter of the $K_6(\alpha)$ function are $k_2 = 4.677$, $k_4 = -0.628$, and $k_6 = 0.071$. It is as expected that the parameter for higher order terms is smaller. The results of MAE dependence on SO coupling strength for LaCo$_5$ and CoPt systems are similar to that of YCo$_5$. The values of fitted $k_2$, $k_4$, and $k_6$ are 4.533, -0.098, and -0.015 for LaCo$_5$ and 1.088, 0.064, and -0.019 for CoPt, respectively. These results together with the deviation of $K_{so}/K$ from 2 discussed above show that the perturbation theory may not be enough to describe the MAE of systems with large MAE. This observation is also consistent with previous works on large MAE systems such as CoPt [31] and FePt [39]. The causes of the deviation of MAE from the perturbation theory are the non-negligible contributions from higher order terms of perturbation theory and/or self-consistent effects [31].

In summary, through a systematic investigation of the dependence of MAE of MCo$_5$ (with M = Y and La) on the Co on-site Coulomb U and exchange J within the GGA + U scheme, we determine a *universal* set of U and J parameters for Co 3$d$-electrons. The results from calculations with this set of parameters are consistent with experiments and previous calculations not only for MCo$_5$ systems but also for CoPt system, which has a very different crystalline structure than MCo$_5$. We show that the key factor for large MAEs of MCo$_5$ systems is the large OMA and not the amplitude of the orbital moment itself. The second order perturbation theory is able to capture the main contributions, but the higher order terms need to be included for a better description of systems with large MAEs. Using the simplified version of the virial theorem with SO interaction, we analyze the nature of the SO interaction screening in solids with Coulomb correlations included.

## ACKNOWLEDGEMENTS

This work is supported by the U.S. Department of Energy (DOE), Office of Science, Basic Energy Sciences, Materials Science and Engineering Division, including the grant of computer time at the National Energy Research Scientific Computing Center (NERSC) in Berkeley, CA. M. C. N. is also supported by U.S. DOE, Energy Efficiency and Renewable Energy, Vehicles Technology Office, EDT program. Work of V. P. A. is also supported by the Critical Materials Institute, an Energy Innovation Hub funded by the U.S. DOE. The research was performed at Ames Laboratory, which is operated for the U.S. DOE by Iowa State University under contract # DE-AC02-07CH11358.
*Email: mcnguyen@ameslab.gov

## REFERNCES